\documentclass{aa}
\usepackage{graphicx}
\usepackage{multirow}
\usepackage{txfonts}
\usepackage{floatflt}
\usepackage[authoryear]{natbib}
\bibpunct{(}{)}{;}{a}{}{,}

\begin{document}
\title{V1309 Scorpii: merger of a contact binary
 \thanks{Based on observations obtained with the 1.3-m Warsaw telescope at
the Las Campanas Observatory of the Carnegie Institution of Washington.
}}
\author{R. Tylenda\inst{1}
\and M. Hajduk\inst{1}
\and T. Kami{\'n}ski\inst{1}
\and A. Udalski\inst{2,3}
\and I. Soszy{\'n}ski\inst{2,3}
\and M. K. Szyma{\'n}ski\inst{2,3}
\and M.~Kubiak\inst{2,3} 
\and G.~Pietrzy{\'n}ski\inst{2,3,4}
\and R.~Poleski\inst{2,3}
\and {\L}. Wyrzykowski\inst{3,5} 
\and K. Ulaczyk\inst{2,3}
}

\offprints{R. Tylenda}
\institute{Department for Astrophysics, N.~Copernicus
            Astronomical Center, Rabia\'nska~8, 87-100~Toru\'n, Poland\\
            \email{tylenda,cinek,tomkam@ncac.torun.pl}
     \and Warsaw University Observatory, Al.~Ujazdowskie~4, 
          00-478~Warsaw, Poland\\ 
          \email{udalski,soszynsk,msz,mk,pietrzyn,rpoleski,kulaczyk@astrouw.edu.pl}
    \and  The Optical Gravitational Lensing Experiment
    \and Universidad de Concepci{\'o}n, Departamento de Astronomia,
         Casilla 160--C, Concepci{\'o}n, Chile
    \and  Institute of Astronomy, University of Cambridge, Madingley Road,
          Cambridge CB3~0HA,~UK \\
          \email{wyrzykow@ast.cam.ac.uk}
}
\date{Received; accepted}
\abstract
{Stellar mergers are expected to take place in numerous circumstences in
the evolution of stellar systems. In particular, they are considered as
a plausible origin of stellar eruptions of the V838~Mon type. V1309~Sco is the most recent
eruption of this type in our Galaxy. The object was discovered in September~2008.}
{Our aim is to investigate the nature of V1309~Sco.}
{V1309~Sco has been photometrically observed in course of the OGLE project
since August~2001. We analyse these observations in different ways. In
particular, periodogram analyses were done to investigate the nature of the
observed short-term variability of the progenitor.}
{We find that the progenitor of V1309~Sco was a contact binary with an
orbital period of $\sim$1.4~day. This period was decreasing with time.
The light curve of the binary was also evolving, indicating that
the system evolved towards its merger. The violent phase of the merger, 
marked by the systematic brightenning of the object,
began in March~2008, i.e. half a year before the outburst
discovery. We also investigate the observations of V1309~Sco during the
outburst and the decline and show that they can be fully accounted for
within the merger hypothesis.}
{For the first time in the literature we show from direct observations 
that contact binaries indeed end up
by merging into a single object, as was suggested in
numerous theoretical studies of these systems. Our study also 
shows that stellar mergers indeed result in eruptions of the V838~Mon type.}


\titlerunning{V1309 Sco}
\authorrunning{Tylenda et al.}
\maketitle
\section{Introduction  \label{intro}}

Stellar mergers have for a long time been recognized to play an important
role in the evolution of stellar systems. 
High stellar densities in globular clusters can lead often to collisions and mergers 
of stars \citep{leon89}. In this way the origin of blue strugglers can be
explained.
In dense cores of young clusters multiple mergers of protostars have been suggested as 
a way of formation of the most massive stars \citep{bonn98}. Some binary stars, 
in particular contact binaries, are suggested to end their evolution as stellar mergers
\citep{robeggl}. 

\begin{figure*}
  \includegraphics[height=\hsize]{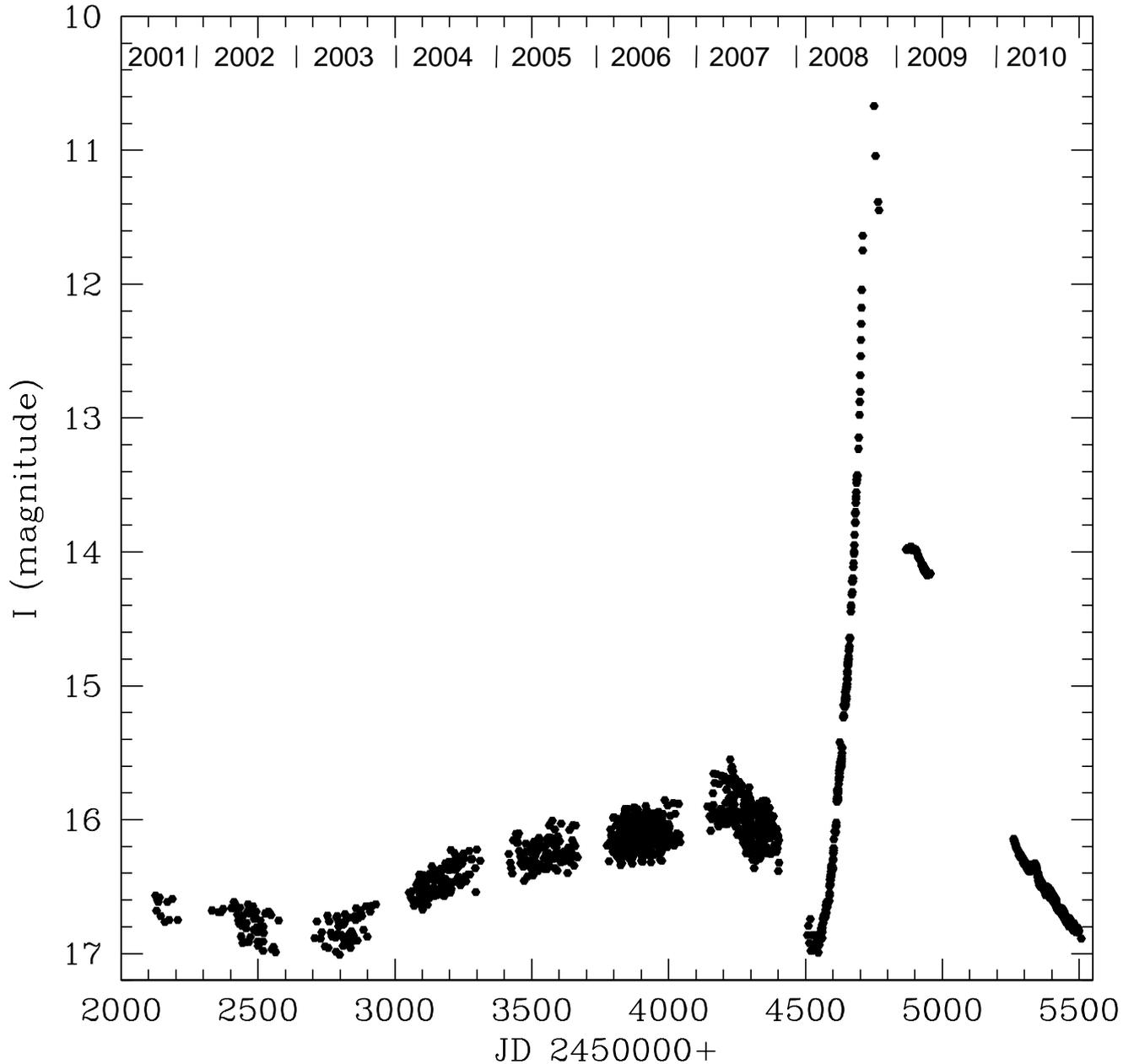}
  \caption{Light curve of V1309 Sco from the OGLE-III
and OGLE-IV projects: $I$ magnitude versus
time of observations in Julian Dates. Time in years is marked on top of
the figure. At maximum the object attained $I \simeq 6.8$.
}
  \label{lightcurve}
\end{figure*}

The powerful outburst of V838 Mon in 2002 \citep{mun02}, 
accompanied by a spectacular light echo \citep{bond03},
raised interest in a class of stellar eruptions named "red novae", "optical transients"
or "V838~Mon type eruptions". 
These objects, which typically reach a maximum luminosity of 
$\sim10^6~{\rm L}_\odot$,
evolve to low effective temperatures and decline as very cool (super)giants.
 Apart form V838~Mon, in our Galaxy the class also includes V4332~Sgr, 
whose outburst was observed in 1994 \citep{martini}, and
V1309~Sco, which erupted in 2008 \citep{mason10}. As extragalactic eruptions
of this kind one can mention M31~RV \citep[eruption in 1989,][]{mould},
M85~OT2006 \citep{kulk07}, and NGC300~OT2008 \citep{berger09}.
Several interpretations of the eruptions were proposed. They include an unusual
classical nova \citep{tutu,shara}, a late He-shell flash \citep{lawlor},
or a thermonuclear shell flash in an evolved massive star \citep{muna05}.
\citet{tylsok06} presented numerous arguments against these mechanisms.
They showed that all the main observational characteristics of the
V838~Mon-type eruptions can be consistently understood as resulting 
from stellar collisions and mergers, as originally proposed in
\citet{soktyl03}. For these reasons \citet{st07} proposed to call these type
of eruptions {\it mergebursts}. Recently, \citet{kfs10} and \citet{ks11}
have suggested that some of the V838~Mon-type
eruptions are of the same nature as the eruptions of luminous blue
variables,
and that they can be powered by mass-transfer events in binary systems. 

In the present paper, we show that the recent red nova, V1309~Sco, 
is a Rosetta stone in the studies of the nature of the V838~Mon type
eruptions. Archive photometric data collected for the object in
the OGLE project during about six years before the outburst allow us to conclude that 
the progenitor of V1309~Sco was a contact binary. The system quickly evolved
towards its merger, which resulted in the eruption observed in 2008.

\section{V1309 Scorpii}
V1309 Sco, also known as Nova Sco 2008, was discovered on 2.5~September~2008
(JD~2454712) \citep{nak08}. 
The subsequent evolution, however, showed that as pointed out in \citet{mason10},  
this was not a typical classical nova. 

 Early spectroscopy revealed an F-type giant \citep{ruda}. On a time
scale of a month the object evolved to K- and early M-types \citep{mason10,rudy}.
Eight months after the discovery it was observed as a late M-type giant \citep{mason10}.

As described in \citet{mason10}, the object developed a complex and rapidly evolving 
line spectrum of neutral and singly ionized elements. In early epochs,
absorption features dominated the spectrum, emission components developed
later and their intensities quickly increased with time. The strongest
emission lines were those of the hydrogen Balmer series.

The emission components had FWHM of $\sim$150~km\,s$^{-1}$ and broader
wings, which in the case of H$\alpha$ extended even beyond 1000~km\,s$^{-1}$
\citep{mason10}. Narrow absorption components were superimposed on
the emission components, so that the line profiles mimicked those of P-Cyg
type or inverse P-Cyg in some cases. \citet{mason10} interpret these line profiles 
as produced in an expanding shell that is denser in the equatorial plane.

In late epochs, when the object evolved
to M-type, absorption features of TiO, VO, CO, and H$_2$O appeared in the
spectrum \citep{mason10,rudy}.

V1309 Sco thus shares the principal 
characteristic of the V838~Mon type eruptions, i.e evolution to very low
effective tempratures after maximum brightness and during the decline \citep{tylsok06}.
Other common features of V1309~Sco and the V838~Mon type eruptions include:
outburst time scale of the order of months, outburst amplitude (maximum
minus progenitor) of 7 -- 10 magnitudes, complete lack of any
high-ionization features (coronal lines, in particular),
expansion velocities of a few hundred km~s$^{-1}$ (instead of a few
thousands as in classical novae), and oxide bands observed in later epochs,
which imply that 
oxygen-rich (C/O~$<$~1) matter was involved in the eruption.

\section{Observations}
Owing to the position of V1309 Sco close to the Galactic centre 
(l = 359\fdg8, b = --3\fdg1),
the object appears to be situated within a field monitored in the OGLE-III
and OGLE-IV projects \citep{udal03}\footnote{http://ogle.astrouw.edu.pl}.
As a result V1309~Sco was observed  on numerous occasions since 
August~2001. Altogether more than 2000 measurements
were obtained predominantly in the $I$ Cousins photometric band.
Among them, $\sim$1340 observations were made before the
discovery of the object as a nova in September~2008. 
A few observations were also made in the $V$ band (seven before the discovery).
The data were reduced and calibrated using standard OGLE procedures
\citep{udal08}.
A typical precision of the measurements was 0.01 magnitude. 

\section{Results \label{result}}
The entire measurements of V1309~Sco derived from the OGLE-III and IV surveys 
in the $I$ photometric band are displayed in Fig.~\ref{lightcurve}. 
The gaps in the data are owing to conjunctions
of the object with the Sun. Apart from 2001 and 2009 most of the data were
obtained between February and October of each year. In 2001 the OGLE-III
project was just starting to operate, hence only 11 measurements were
obtained in this year (they were omitted from the analysis discribed below).
In the period 2002--2008, from 52 (2002, 2003) to 367 (2006) observations 
were made each year. 
Near the maximum of the 2008 eruption, i.e. when the object was
brighter than $I \simeq 11$, its image was overexposed in the OGLE
frames, hence there are no data for this period.  Near maximum brightness,
the object attained $I \simeq 6.8$, according to the data gathered in 
AAVSO\footnote{available at http://www.aavso.org/}.
In May~2009 OGLE-III phase ended and threfore only 64 data points were
collected during the 2009 observing season. The OGLE project resumed regular  
observations of the Galactic centre in March~2010 with a new
instrumental setup, namely a 32 chip mosaic camera, which started the OGLE-IV
phase. In 2010 655 measurements were obtained. 

As can be seen from Fig.~\ref{lightcurve},
the progenitor of V1309~Sco was initially 
slowly increasing in brightness on a time scale of years and reached a local maximum 
in April 2007. Subsequently, the star faded by $\sim$1 magnitude during a year. 
In March 2008, the main eruption started, which led to the object discovery, 
as Nova Sco 2008, half a year later. Remarkable is the smooth, roughly exponential 
rise in brightness. During this event the object brightened by $\sim$10 magnitudes, i.e. 
by a factor of $10^4$. An analysis of the data for the progenitor (2002--2007)
is presented in Sect.~\ref{progen}, while the outburst and the decline of
the object are discussed in Sect.~\ref{sect_burst}.

\section{The progenitor  \label{progen}}

\subsection{The data  \label{prog_dat}}
\begin{figure}
  \includegraphics[height=\hsize]{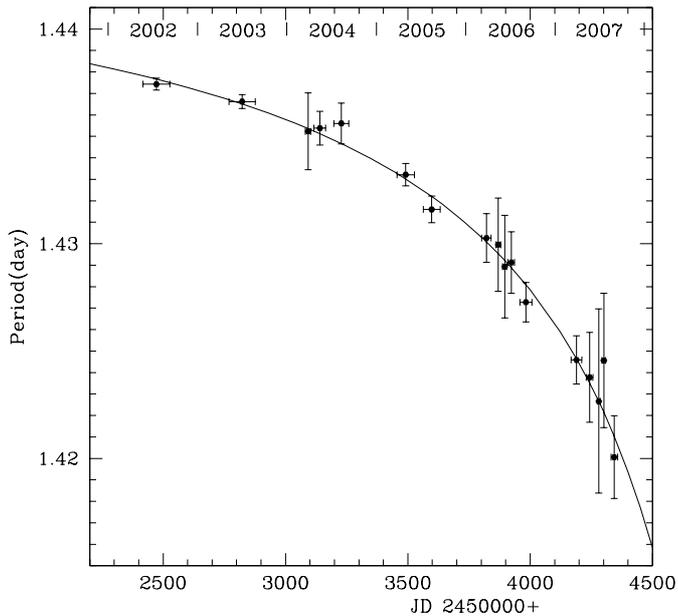}
  \caption{Evolution of the period of the photometric variations of the
V1309~Sco progenitor.
The line shows a least-squares fit of an exponential formula to the data (see
text and Eq.~\ref{per_fit}).}
  \label{fig_period}
\end{figure}

A short-term variability,
resulting in a $\sim$0.5 magnitude scatter of the points in
Fig.~\ref{lightcurve}, is the most remarkable and interesting feature of 
the V1309~Sco progenitor. This variability is
strictly periodic. To show this, we used 
a method that employs periodic orthogonal polynomials to fit the observations and
an analysis of variance to evaluate the quality of the fit,
as described in \citet{schwarz}. This method is particularly suitable for
analysing unevenly sampled data of non-sinusoidal periodic variations, 
such as eclipses in binary systems. The resulting periodograms obtained for
the particular observational seasons are presented in Appendix~\ref{periodograms}. 
They show that the
principal periodicity in the variations observed in seasons 2002 -- 2007
corresponds to a period of $\sim$1.4 day. 
However, the derived period was not constant, but slowly decreasing with time, 
as presented in Fig.~\ref{fig_period}. 
It decreased by 1.2\% during the pre-outburst observations. 

When deriving the period plotted in Fig.~\ref{fig_period},
 we devided the data sequence available for a given season 
into subsamples, each containing $\sim$50 data points, 
and constructed periodograms for each subsample separately. In this way we
can see that during later seasons, particularly in 2006 and 2007, the period
was significantly varying on a time scale of months.

We find that the time evolution of the observed period can be fitted 
by an exponential formula. A result of a least-squares fit (inverse errors
used as weights) is shown with the full line in Fig.~\ref{fig_period}. The
obtained formula is
\begin{equation}
P = 1.4456\ {\rm exp}\ \Big( \frac{15.29}{t - t_0} \Big),
\label{per_fit}
\end{equation} 
where $P$ is the period in days, 
$t$ is the time of observations in Julian Dates, and $t_0$ = 2455233.5. 

\begin{figure}
  \includegraphics[height=\hsize]{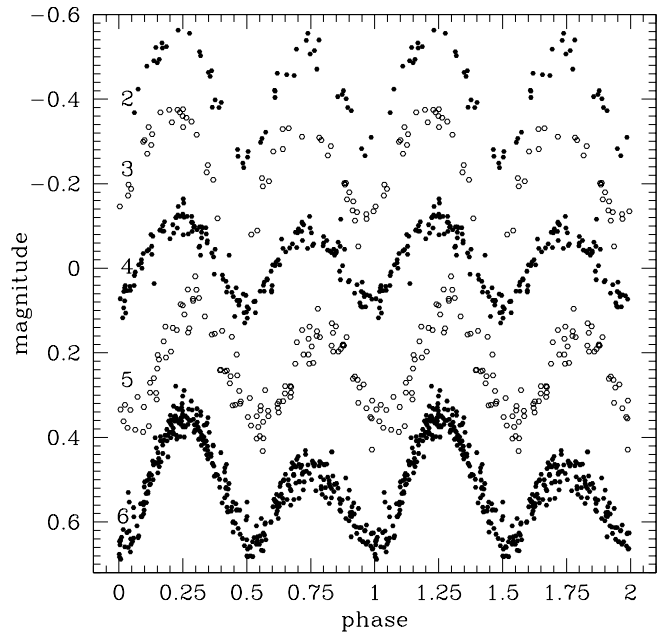}
  \includegraphics[height=\hsize]{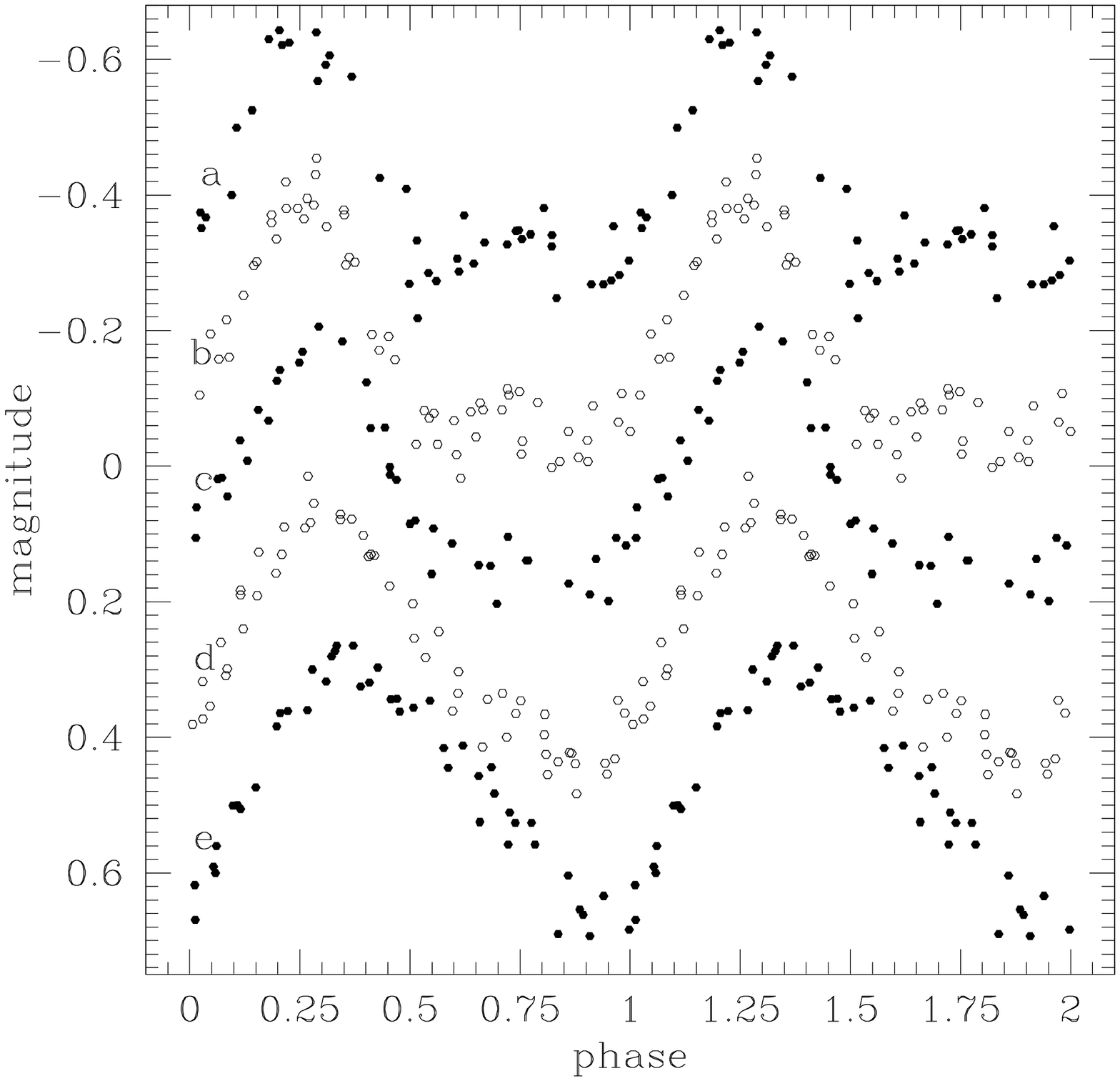}
  \caption{Light curves obtained from folding the data with the period
  described by Eq.~(\ref{per_fit}). Upper part: seasons 2002 -- 2006. Lower part:
  season 2007 devided into five subsamples (time goes from a to e). The zero
  point of the magnitude (ordinate) scale is arbitrary.}
  \label{fig_lc}
\end{figure}

Using the above period fit 
we folded the observations and obtained light curves of the object in 
particular seasons. 
The results are displayed in Fig.~\ref{fig_lc}. 
In 2002 -- 2006 the light curves were obtained
from all available data for a particular seasons. The
results are shown in the upper part of Fig.~\ref{fig_lc}. In 2007
the light curve significantly evolved during the season and therefore
 we plotted the results for the
subsamples (the same as when deriving the period) of 2007 season
in the lower part of Fig.~\ref{fig_lc}
(time goes from subsample a to e in the figure).

As can be seen from Fig.~\ref{fig_lc}, the light curve in 2002 -- 2006
displays two maxima and two minima. They are practically
equally spaced in the phase and have similar shapes in the early seasons. 
This is the reason why
the periodograms for these seasons (see Appendix~\ref{periodograms}) show a
strong peak at a frequency twice as high (a period twice as short) as the
main peak. In other words, the observations could have been interpreted with a
period of $\sim$0.7~day and a light curve with one maximum and minimum. The
periodograms show however already in the 2002 season that the period that is 
twice as long
($\sim$1.4~day) better reproduces the observations. The reason is obvious in
the later seasons, when the first maximum (at phase 0.25 in Fig.~\ref{fig_lc})
in the light curve becomes increasingly stronger than the second one. As a
result the peak corresponding to the 0.7~day period deacreases and 
practically disappears in 2007.

\subsection{Interpretation  \label{interpret}}
In principle one can consider three possible interpretations of 
the observed light variability, 
i.e. stellar pulsation, single-star rotation, and an eclipsing binary system.

\subsubsection{Pulsation}
The light curve in the early seasons may imply a stellar pulsation with a period of
$\sim$0.7~day 
and an amplitude of $\sim$0.15 magnitude. This interpretation, however, encounters severe 
problems when it is used to explain the observed evolution of V1309 Sco. As we show below, 
the progenitor was probably a K-type star. There is no class of pulsating stars
that could be reconciled with these characteristics, i.e a K-type star pulsating with 
the above period and amplitude. Moreover, to explain the observed evolution of 
the light curve displayed in Fig.~\ref{fig_lc}, one would have to postulate a switch of 
the star pulsation to a period exactly twice as long on a time scale of a few years, 
together with a gradual shortening of the period. This would be very difficult, 
if not impossible, to understand within the present theory of stellar pulsation. 
Finally, there is no physical mechanism involving or resulting from a pulsation instability, 
which could explain the powerful 2008 eruption, when the object brightened by a factor of
$10^4$.

\subsubsection{Rotation of a single star}
The observed light curve could have been explained by a single K-type star rotating with 
a period of $\sim$1.4 day and with two spots (or rather two groups of spots) 
on its surface in early seasons. 
The two spots would be replaced by one spot in 2007.  The object would thus belong to 
the FK Com class of rapidly rotating giants \citep[e.g.][]{bopp81}. 
There are, however, several differences 
between the progenitor of V1309 Sco and the FK Com stars. The 1.4 day period is
short, 
as for the FK Com stars. The very good phase stability of the observed light curve, 
shown in Fig.~\ref{fig_lc}, implies that the spot(s) would have had to keep the same position(s) 
on the star surface over a time span of a few years. This seems to be very improbable 
given differential rotation and meridional circulation, which are expected to be substantial 
for a rapidly rotating star. Indeed, for the FK Com stars, 
the spots are observed to migrate and change their position on much shorter time scales
\citep{korh07}.
The systematic decrease of the rotational period is also difficult 
to explain for a single star, and is not observed in the FK Com stars. 
Finally, no eruption such as that of V1309 Sco in 2008 was observed for the FK Com stars. 
Indeed, there is no known mechanism that could produce such a huge eruption in 
a single giant, even if it were fast rotating.

\subsubsection{A contact binary evolving to its merger  \label{sect_cb}}
We are thus left with an eclipsing binary system. As we show below, this possibility 
allows us to explain all principal characteristics of the observed evolution of 
the V1309 Sco progenitor, as well as the 2008 outburst. 

The shape of the light curve, 
especially in early seasons (two rounded maxima of comparable brightness and two equally 
spaced minima), implies that the progenitor of V1309 Sco was a contact binary. 
The orbital period of $\sim$1.4 day does not allow us to classify the object
as a W~UMa-type binary, because the clasical W~UMa stars have periods
$<1$~day. Nevertheless, contact binaries
with periods beyond 1~day are also observed \citep{pacz06}. The exponentially decreasing 
period (Fig.~\ref{fig_period}) can be interpreted as resulting from an unstable phase of 
evolution of the system, which leads to the shrinkage of the binary orbit and 
finally to the merger of the system. 
Such a situation was predicted in theoretical studies
 \citep{webb76,robeggl,rasio95} and it is now commonly believed 
that the W~UMa binaries end their evolution by merging into a single star. Dissipation 
of the orbital energy in the initial, violent phase of the merger
\citep{soktyl06} resulted in the 
V1309 Sco eruption observed in 2008.

One of the possible ways that can lead a binary system to merge is
the so-called Darwin instability.
This happens when the spin angular momentum of the system is more than a
third of the orbital angular momentum. As a result,
tidal interactions in the system cannot maintain the primary 
component in synchronization anymore. The orbital angular velocity is higher than 
the primary's angular velocity, the tidal forces increase their action and rapidly transport 
angular momentum from the orbital motion to the primary's rotation. For
contact binaries this takes place when the binary mass ratio 
$q \equiv M_2/M_1 \la 0.1$ \citep{rasio95}. 

Another possibility appears to be if a binary system enters a deep contact,
as discussed in \citet{webb76}. This can happen, for instance, if the primary
attempts to cross the Hertzsprung gap. The system then starts loosing mass
and angular momentum through the outer Lagrangian point, $L_2$. This shrinks
the binary, which further deepens the contact and increases mass and angular
momentum loss. In addition
the system starts orbiting faster than the components rotate. As a result,
similar as for the Darwin instability, the tidal forces transport
angular momentum from the orbit to the components' spins, which further accelerates
the orbit shrinkage.

A maximum in the light curve of a typical contact binary is observed when we look 
at the system more or less perpendicularly to the line that joins 
the two components. As a result, we observe two maxima during each orbital revolution. 
The maxima are of similar brightness, because the system looks similar from both sides. 
The situation changes when one of the above instabilities sets in. The secondary 
starts orbiting faster than the primary's envelope rotates. The stars are in contact, 
which means that the difference in velocities is partly dissipated near the contact 
between the components. 
This should lead to the formation of a brighter (hotter) region that is
visible when we look at the 
leading side of the secondary. The system is then brighter than when we look from 
the opposite direction. The maxima in the light curve start to differ. This
is what we observe in the case of the V1309 Sco progenitor. As can be seen from
Fig.~\ref{fig_lc}, 
the first maximum (at phase $\sim$0.25) becomes progressively stronger than the second one 
(phase $\sim$0.75) from 2002 to 2006. In 2007, the second maximum 
disappears, and at the end of this season we observe a light curve with only one maximum 
and one minimum. Apparently the system evolved to a more spherical configuration with 
a bright (hot) spot covering a large fraction of the system's surface along 
the orbital plane.

\subsection{Basic parameters of the progenitor}

\subsubsection{Interstellar reddening, spectral type, and effective temperature
              \label{sect_redd}}
\begin{figure}
  \includegraphics[height=\hsize]{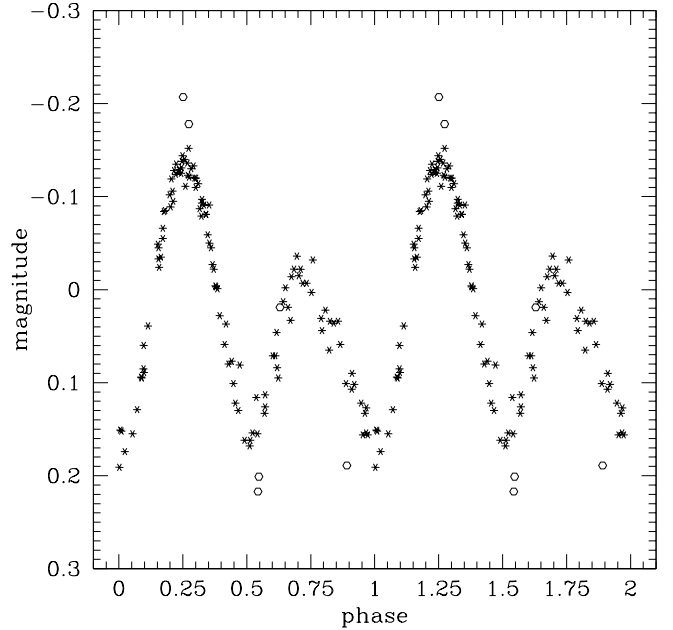}
  \caption{$V$ (open points) and $I$ (asterisks) measurments obtained in
JD~2453880 -- 2453910 (season 2006) and folded with the period
  described by Eq.~(\ref{per_fit}). The zero point of the magnitude
(ordinate) scale is arbitrary.
}
  \label{fig_VI}
\end{figure}

The effective temperature of the progenitor can be derived from the 
$V - I$ colour.
In 2006 a few measurements were obtained in the $V$ band. They are
displayed in Fig.~\ref{fig_VI} together with the $I$ data obtained in the
same time period (JD~2453880 -- 2453910).

In order to determine the effective temeprature from a colour, an estimate of the 
interstellar reddening is also necessary. \citet{mason10} obtained 
$E_{B-V} \simeq 0.55$ from 
interstellar lines observed in their spectra of V1309 Sco in outburst. We constrained 
spectral types of the object in outburst from the UVES spectra of \citet{mason10} 
obtained in their epochs 1 -- 3 (downloaded from the VLT archive). 
Comparing them with the $BVRI$ photometry of the object obtained by the AAVSO team 
near the above spectral observations, we obtained $0.7 \la  E_{B-V} \la 0.9$. 
The OGLE data also 
allowed us do derive a $V - I$ colour, equal to 5.27, 223 days after the outburst 
discovery date. 
This happened to be in the epoch of the SOAR spectroscopy of \citet{mason10}, from which 
these authors derived a spectral type of M6--7. To reconcile the colour with the spectral 
type, $E_{B-V} > 0.7$ is required.  An upper limit to the extinction can be
obtained from the Galactic Dust Extinction
Service\footnote{http://irsa.ipac.caltech.edu/applications/DUST/},
which estimates $E_{B-V}$ from the 100~$\mu$m dust
emission mapped by IRAS and COBE/DIRBE. 
For the position of V1309~Sco, we obtain $E_{B-V} \la 1.26\pm0.11$.
Below we assume that V1309 Sco is reddened with 
$E_{B-V} \simeq 0.8$. 

As can be seen from Fig.~\ref{fig_VI}, two $V$ measurements were taken
close to the first maximum in the light curve (phase $\sim$0.26), two others near the 
following minimum (phase $\sim$0.55). The resulting colour, $V - I$, was 2.09 and 2.21 for 
the two phases, respectively. 
The mean value of the observed colour, i.e. $V - I \simeq 2.15$, corrected for 
the reddening and compared with the standard colours of giants (luminosity class III) 
results in a spectral class K1--2 and an effective temperature 
$T_{\rm eff} \simeq 4500$~K for the progenitor of V1309 Sco.

The difference in the colour, as mentiond above, implies that at the maximum the object was
on average $\sim$200~K hotter than in the minimum. 
This agrees with the interpretation of 
the evolution of the light curve given in Sect.~\ref{sect_cb}.

\subsubsection{Luminosity and distance  \label{sect_ld}}
The hypothesis that the progenitor of V1309 Sco was a contact binary allows us 
to evaluate other parameters of the system. Adopting a total mass of the binary, 
$(M_1 + M_2)$, of 1.0 -- 3.0~M$_\odot$ (observed range of masses of W~UMa binaries) and 
an orbital period of 1.43~day, one obtains from the third Kepler law that 
the separation of the components, $A$,  is $(3.7 - 5.4)\times 10^{11}$~cm
(5.4 -- 7.7~R$_\odot$). 

Taking the mass ratio, $q$,  within the range observed in the W~UMa systems,
i.e. 0.07 -- 1.0,
we can derive the radii of the Roche lobes of the components \citep{eggl83}.
Assuming that both components fill their Roche lobes a total effective
surface of the system can be obtained.
This, along with the above estimate of $T_{\rm eff}$ results in a luminosity of the
system of $(1.15 - 3.3) \times 10^{34}$~erg~s$^{-1}$ (3.0 -- 8.6~L$_\odot$).

Taking into account bolometric correction, 
interstellar reddening, and the mean observed brightness of the V1309 Sco progenitor 
in 2006, $V = 18.22$, we derive a distance to the object of 
$3.0 \pm 0.7$~kpc. If the outer Roche lobes \citep[dimensions can be calculated
from][]{yakegg} are assumed to be filled by the
components, the luminosity increases by $\sim$35\%, while the distance
becomes $3.5 \pm 0.7$~kpc. The
distance estimate is almost independent of the reddening value.

The above estimates of the 
luminosity and effective temperature are consistent with a $\sim$1.0~M$_\odot$ star 
at the beginning of the red giant branch
when they are compared with theoretical tracks of stellar 
evolution \citep[e.g.][]{girardi}, 

\section{The outburst  \label{sect_burst}}

\subsection{The data \label{burst_data}}
\begin{figure}
  \includegraphics[scale=0.4]{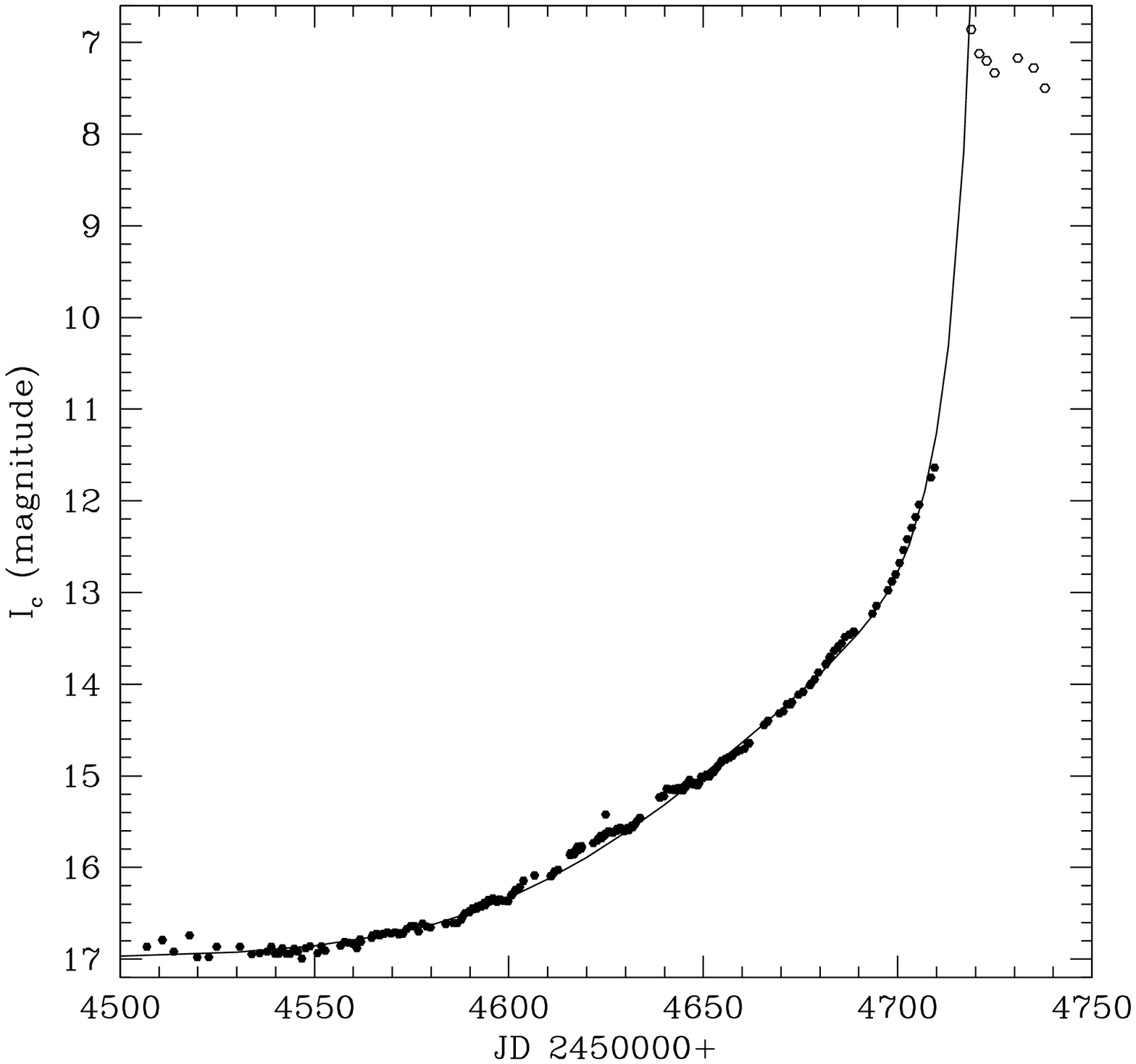}
  \includegraphics[scale=0.4]{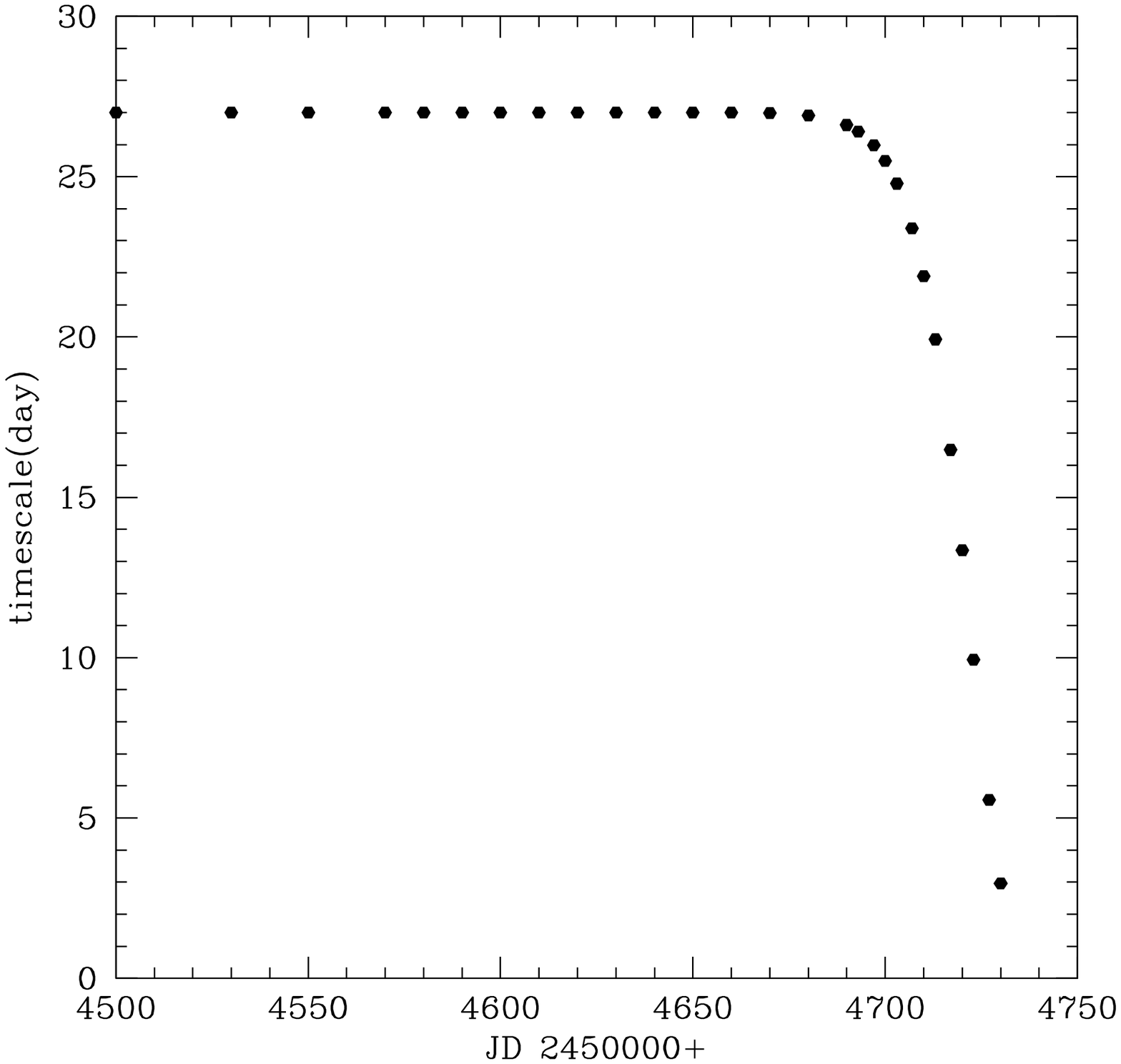}
  \caption{$I$ light curve of V1309~Sco during its rise to maximum in
2008 (upper part). Full points: data from OGLE~III, open points: data from AAVSO. 
The line shows a fit of an exponential formula (Eq.~\ref{eq_rise}). The time scale
used in the formula
is plotted in the lower part of the figure.}
  \label{fig_rise}
\end{figure}

Figure~\ref{fig_rise} (upper part) displays the rise of V1309~Sco 
during its eruption in 2008 (open points near maximum show the $I$ results
taken from AAVSO\footnote{available at http://www.aavso.org/}).
This phase started in March~2008, and it took the object about six months to
reach the maximum brightness. The rise was remarkably smooth for an
eruption that led the object to brighten by a factor of $\sim 10^4$.
We found that the light curve can be fitted by an exponential formula, 
i.e.
\begin{equation}
  \label{eq_rise}
 I = -2.5\ {\rm log} (F_I),
 ~~~~~~~F_I = F_0 + F_1\ {\rm exp} (t/\tau(t)),
\end{equation}
where $F_0 = 1.6 \times 10^{-7}$, $F_1 = 2.1 \times 10^{-8}$ are fluxes in
the Vega units, $t$ = JD -- 2454550, and $\tau(t)$ is a time scale, allowed
to evolve with time. The time evolution of $\tau$, resulting in the fit 
shown with the full line in the upper part of Fig.~\ref{fig_rise},
is displayed in the lower part of the figure.
These results show that during the
initial $\sim$5~months of the rise, 
i.e. up to JD~$\simeq$~2454690, the time scale
was constant and equal to $\sim$27~days. During this phase
the object brightened by $\sim$3.5~magnitudes. Then the time scale quickly
decreased and the object brightened by $\sim$6.5~magnitudes during a month.

\begin{figure}
  \includegraphics[height=\hsize]{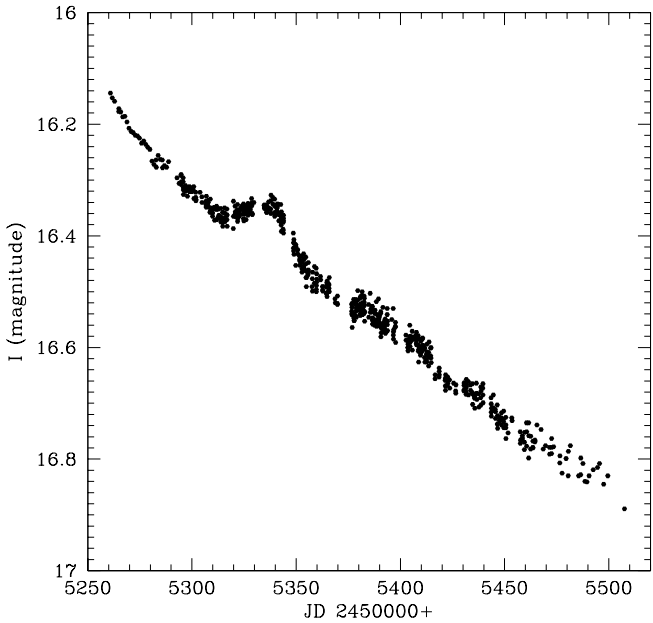}
  \caption{$I$ light curve of V1309~Sco during the decline in 2010.}
  \label{fig_decl}
\end{figure}

The $I$ light curve of V1309~Sco during the decline observed by OGLE-IV in 2010 
is presented in Fig.~\ref{fig_decl}. Evidently the decline was relatively smooth 
at the begining of the presented phase. A similar
behaviour (smooth evolution) was also observed in 2009 (see
Fig.~\ref{lightcurve}). However, as the decline continues in 2010,
oscillations of a few hundredths magnitude on a time scale of hours appear. 
A periodogram analysis
shows that there is no significant periodicity in these light variations.

In a few cases multiband photometric
observations are available during the outburst and decline. 
They allow us to estimate the spectral type and
effective temperature of the object.
This is the case for seven dates in September~2008, when multiband photometry
was made by the AAVSO team.
On 4 and 16~October~2008 the object was observed by \citet{rudy} using the
Infrared Telescope Facility, and $JHK$ magnitudes were derived. These results
can be combined with $BV$ AAVSO measurements and an $I$ OGLE magnitude obtained 
in the same time period.
On 13~April~2009, as well as in a few cases in March--April~2010 and
October~2010, $V$ photometric measurements (together with $I$)  were 
obtained by OGLE. 
On 28~August~2010 the object was observed by one of us
(M.H.) with the SAAO 1.0 m. Elizabeth Telescope with the Bessell $BVRI$ filters. 
The observations were reduced using standard procedures
in the IRAF package. The photometry of the star was obtained with the
DAOPHOT package implemented in IRAF.
The resulting magnitudes are $B = 22.21 \pm 0.10$, $V = 20.42 \pm 0.13$, 
$R = 18.84 \pm 0.10$, and $I = 16.67 \pm 0.10$. The latter
value agrees very well with an OGLE result ($I = 16.69 \pm 0.01$) obtained 
on the same date. 

\begin{table}
\begin{minipage}[t]{\hsize}
\caption{Basic parametres of V1309 Sco in outburst and decline}
\label{tab_evol}
\centering 
\begin{tabular}{cccccc}
\hline
\hline
 JD & data source & sp.type & $T_{\rm eff}$ & $R_{\rm eff}({\rm R}_\odot)$ &
 L (${\rm L}_\odot$) \\[2pt]
\hline
 4718.9 & AAVSO & F9 &  5830 &  174. & $3.16\ 10^4$ \\  
 4721.0 & AAVSO & G1 &  5410 &  177. & $2.40\ 10^4$ \\
 4722.9 & AAVSO & G5 &  4870 &  209. & $2.22\ 10^4$ \\
 4724.9 & AAVSO & K1 &  4360 &  252. & $2.06\ 10^4$ \\
 4730.9 & AAVSO & K3 &  4150 &  297. & $2.36\ 10^4$ \\
 4735.0 & AAVSO & K4 &  3980 &  312. & $2.21\ 10^4$ \\
 4737.9 & AAVSO & K5 &  3900 &  297. & $1.84\ 10^4$ \\
 4750.  & AAVSO/ & M4 &  3510 &  155. & $3.27\ 10^3$ \\
        & OGLE/ITF &         &       &       &  \\
 4934.8 & OGLE  & M7 &  3130 &   34. &  98.5 \\
 5282.0 & OGLE  & M5 &  3370 &    9.5&  10.4 \\
 5437.0 & SAAO &  M4 &  3420 &    6.9 &  5.9 \\
 5474.5 & OGLE &  M3 &  3565 &    5.4 &  4.2 \\
\hline
\end{tabular} 
\end{minipage}
\end{table}
  
\begin{figure}
  \includegraphics[scale=0.34]{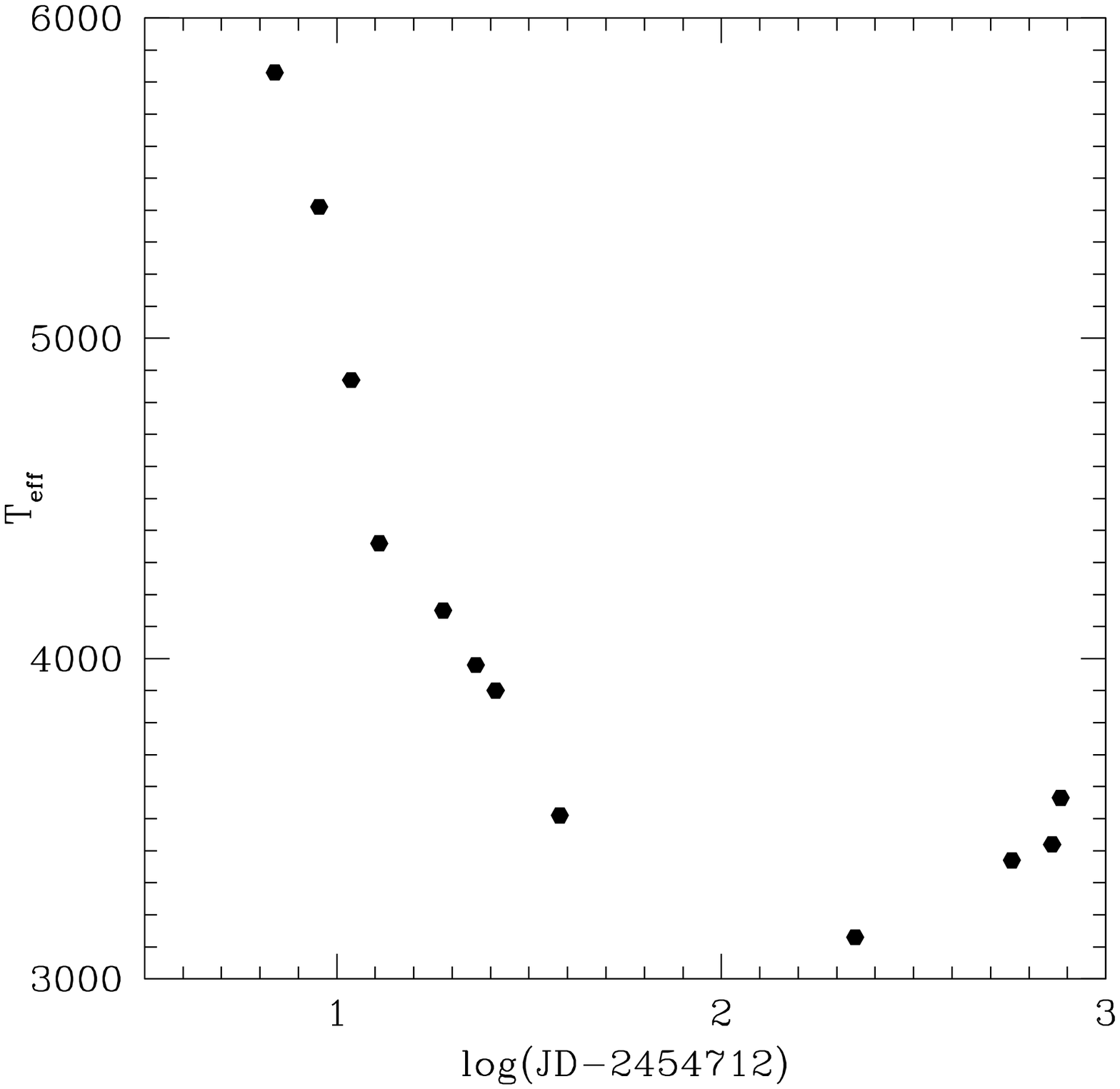}
  \includegraphics[scale=0.34]{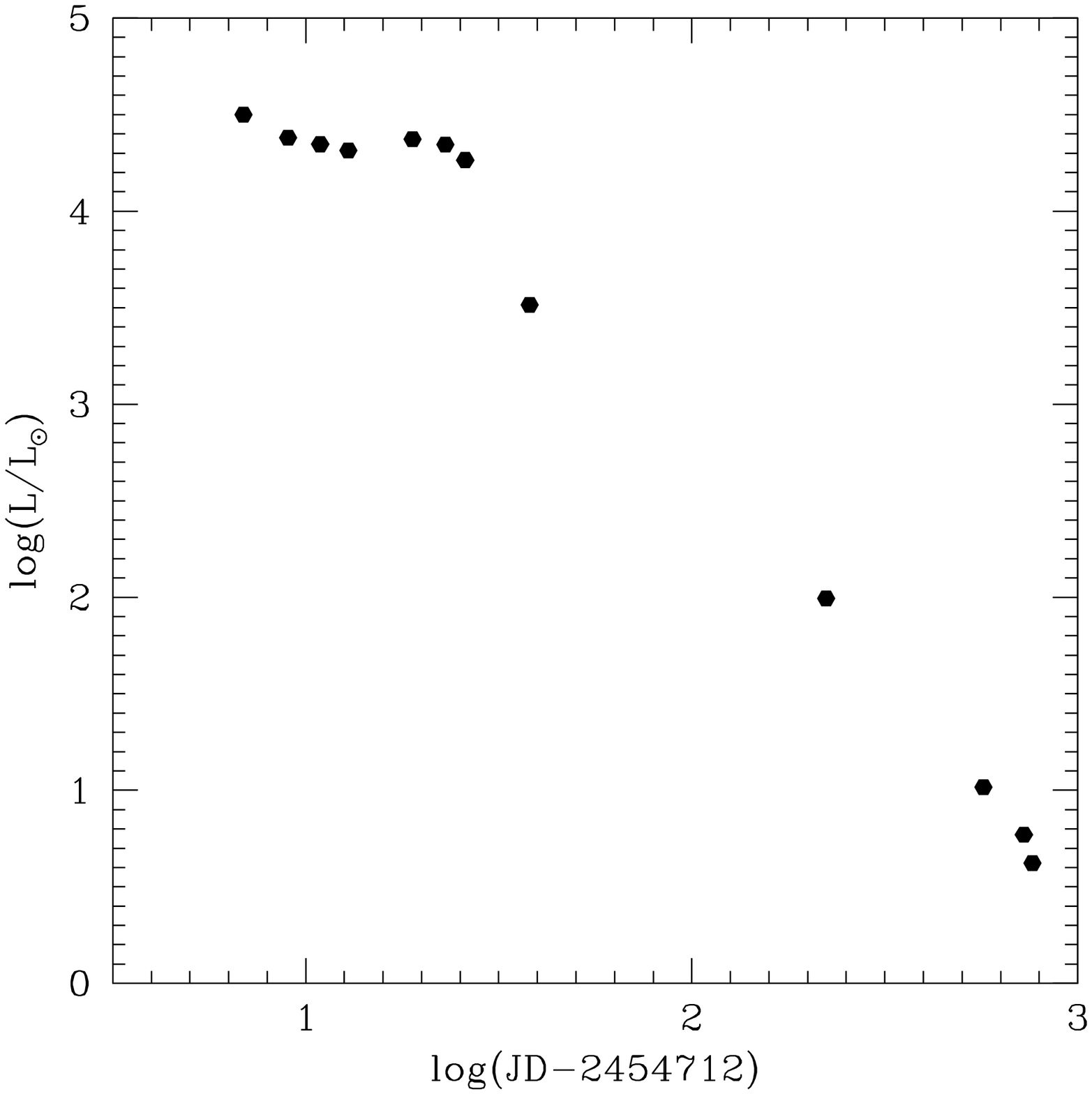}
  \includegraphics[scale=0.34]{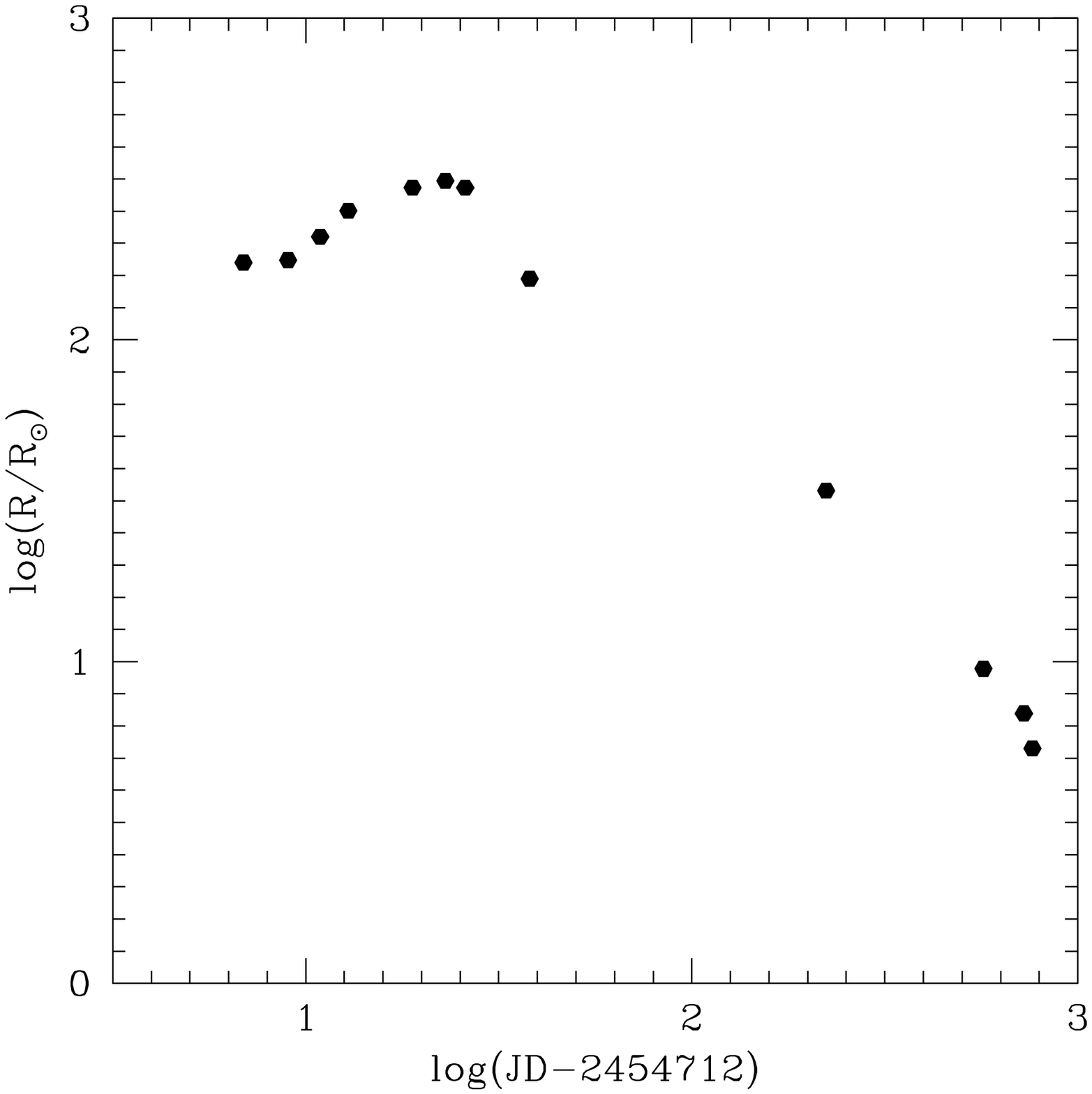}
  \caption{Evolution of V1309 during outburst and decline 
          (see Table~\ref{tab_evol}). The abscissa displays a logarithm of
time in days counted from the date of the discovery.}
  \label{fig_evol}
\end{figure}

\subsection{Observed evolution}
Taking our estimates of the distance (3.0~kpc, see Sect.~\ref{sect_ld}) 
and reddening ($E_{B-V} = 0.8$, see Sect.~\ref{sect_redd}), 
we can derive basic parameters
of V1309~Sco during the outburst from multiband photometry with the same
method as in \citet{tyl05}.
The resulting values of the effective temperature, radius, and
luminosity are presented in Table~\ref{tab_evol} and Fig.~\ref{fig_evol}.
They show that the evolution of V1309~Sco during outburst and decline was
really of the same sort as those of V838~Mon \citep{tyl05} and V4332~Sgr
\citep{tcgs05}. In all these cases the main decline in luminosity was
accompanied by a decline in the effective temperature. In later phases of
the decline the objects resume a slow increase in the effective
temperature.
There is no doubt that V1309~Sco was not a classical nova.
We know from the present study that its eruption resulted from a
merger of a contact binary. Below we show that the energy budget of the
eruption can also be well accounted for by dissipation of the orbital energy of
the progenitor.

\subsection{Merger-powered outburst -- mergerburst}
 The orbital energy and angular momentum of a binary system can be calculated from
\begin{equation}
  \label{eorb_eq}
E_{\rm orb} = -\frac{G\ M_1\ M_2}{2\ A}
            = -\frac{G\ (M_1 + M_2)^2}{2\ A} \frac{q}{(1 + q)^2}
\end{equation}
and
\begin{equation}
  \label{jorb_eq}
J_{\rm orb} = \Big( \frac{G\ A}{M_1 + M_2} \Big)^{1/2} M_1\ M_2 
            = \big( G\ (M_1 + M_2)^3\ A \big)^{1/2}  \frac{q}{(1+q)^2}.
\end{equation}
Taking the values assumed and obtained in Sect.~\ref{sect_ld} 
for the V1309~Sco progenitor, one obtains 
$E_{\rm orb} = (0.2 - 5.6)\ 10^{47}$~ergs and
$J_{\rm orb} = (0.9 - 22.)\ 10^{51}$~g\,cm$^2$\,s$^{-1}$.

As derived in Sect.~\ref{prog_dat},
the orbital period decreased by 1.2\% during $\sim$6 years. 
Thus the system contracted by 0.8\% and its orbital energy decreased by
$\sim 10^{45}$~ergs during this time period, 
which gives a mean rate of energy dissipation of 
$\sim 10^{37}\ {\rm erg\ s}^{-1}$.
Already a small portion from this can account for the observed brightening of the object 
between 2002 and the beginning of the 2007 season (see Fig.~\ref{lightcurve}).

During the same time period the system also lost 0.4\% of its orbital
angular momentum. If the system started shrinking because of the Darwin instability,
the orbital angular momentum loss was primarily absorbed by the spinning-up
primary. If, however, the system entered
a deep contact resulting from evolutionary changes of the components,
mass loss through the $L_2$ point would be the main source of
the orbital angular momentum loss. Assuming that the outflowing mass
carries out a specific angular momentum of $\Omega r_{L_2}^2$, where $\Omega$
is the angular rotational velocity of the system and $r_{L_2}$ is a distance
of $L_2$ from the mass centre, we can estimate that the system lost 
$(0.16 - 2.0) 10^{-3}~M_\odot$ in 2002--2007. This is very much an upper limit
for the mass loss. First, judging from the K spectral type, the system might
have had magnetically active components. Then the outflowing matter might
have kept corotation to radii larger than $r_{L_2}$ and thus have carried out a
specific angular momentum larger than assumed in the above estimate. Second,
matter already lost through $L_2$ may have formed an excretion disc around
the system. Tidal interactions between the binary system and the
disc can transfer angular momentum directly from the system to the disc.
Third, a part of the orbital angular momentum probably went to accelerating
the spins of the components. 

An excretion disc is expected to be formed even if the orbital shrinkage was
initiated by the Darwin instability. Decreasing separation between the
components results in a deeper contact, and mass loss through $L_2$ must finally occur.

Apparently we see in the light curve of the V1309~Sco progenitor
a signature of an excretion disc formation.
We observed the system near the equatorial plane 
(eclipsing binary). Therefore, when the disc became sufficiently massive
and thick, it could result in 
dimming the object for the observer. This is a likely explanation of
the fading of the V1309~Sco progenitor observed during a year before 
the beginning of the 2008 eruption 
(see Fig.~\ref{lightcurve}).

The progressive shortening of the orbital period finally led to engulfing the
secondary by the primary's envelope. This probably took place at some point in
February~2008, when the signs of the binary motion disappeared from the light
curve. The secondary, or rather its core, now spiralling in the
common envelope, started to release the orbital energy
and angular momentum at an increasing rate. The result was the gradual
and relatively gentle brightening of the object, doubling the brightness every
$\sim$19~days (Fig.~\ref{fig_rise}). This phase lasted until $\sim$20~August
(JD $\simeq$2454700), when the eruption abruptly accelarated and the object
brightened by a factor of $\sim$300 during $\sim$10~days. Perhaps this was
a signature of a final disruption of the secondary's core deeply in the
envelope.

In the merger process, especially during its initial, relatively
gentle phases until about August~2008, mass loss probably occured mainly in 
directions close to the orbital plane of the progenitor. As a result, it is
likely that an
extended disc-like envelope was then formed, where a significant
portion of the angular momentum of the progenitor was probably stored. The
main eruption in August~2008, partly blocked by the envelope, was then
likely to occur mainly along the orbital axis.
\citet{mason10} interpret the line profiles observed during the outburst,
especially those of the Balmer series, as produced in a partially collimated
outflow and a slowly expanding shell that is denser in the equatorial plane.

As can be seen from Table~\ref{tab_evol},
V1309 Sco attained a maximum luminosity of $\sim 3\ 10^4~{\rm L}_\odot$. 
The OGLE data show that the period
when V1309~Sco was brighter than $I \simeq 11$~magnitude lasted 40~days.
Both the maximum luminosity and the time scale of the outburst are close 
to the theoretical expectations of a solar-mass star merging with a 
low-mass companion \citep{soktyl06}. 
During $\sim$30~days, when the luminosity of V1309~Sco was 
$\ga 10^4~{\rm L}_\odot$, the object radiated
energy of $\sim 3\ 10^{44}$~ergs. This is $\la$1\% of the energy available in the binary 
progenitor (see above). Of course a considerable energy was also lost during the 
six-months rise, as well as in the decline after maximum. Certainly a significant 
amount was carried out in mass loss. Finally, an energy was also stored in the inflated 
remnant. Estimates made for V838 Mon show that the total energy involved in the outburst 
can be a factor of 10 -- 20 higher than the energy observed in radiation
\citep{tylsok06}.  Even this case can be accounted for by the available orbital energy.

Our SAAO photometry performed in August~2010, as well as the OGLE $V$ and $I$
measurements obtained in 2010 (see Sect.~\ref{burst_data}), 
i.e about two years after the outburst maximum, show that V1309~Sco reached a
luminosity (see Table~\ref{tab_evol}) comparable with its preoutburst value
(see Sect.~\ref{sect_ld}). The object was then significantly 
cooler however than before the outburst (early M-type spectrum versus K1--2). 
The latter agrees with what was
observed in V838~Mon and V4332~Sgr, displaying M-type spectra several years after
their outbursts \citep{tyl05,tcgs05}. However, the drop in luminosity of
V1309~Sco was unexpectedly deep. Both V838~Mon and V4332~Sgr at present
(many years after outburst) remain significantly more luminous than their
progenitors.  Most likely the remnant of V1309~Sco 
contracting after the outburst partly disappeared for us 
behind the disc-like envelope. Moving small-scale blobs of the envelope
matter, absorbing and scattering the light from the central object, could
have been responsible for the short-term variability shown in Fig.~\ref{fig_decl}.
The situation thus looks to be similar to that of the
V4332~Sgr remnant, where the central object is most likely hidden in 
an opaque dusty disc \citep{kst10,kt11}.

\section{Conclusions}
The principal conclusion of our study is that all observed properties of 
V1309~Sco, i.e. the light curve of the progenitor during six years before
the 2008 eruption, as well as the outburst itself, can be consistently explained by 
a merger of a contact binary. This is the first case of a direct
observational evidence showing that 
the contact binary systems indeed end their evolution by merging into single objects,
as predicted in numerous theoretical studies of these systems.

Our study also provides a conclusive evidence in favour of the hypothesis that 
the V838~Mon-type eruptions (red novae) result from 
stellar mergers, as originally proposed in \citet{soktyl03} and
\citet{tylsok06}. In particular, long- and short-term variabilities of 
the progenitors, as those of V838~Mon and V4332~Sgr \citep{goran07,kimes07},
which were sometimes raised as evidence against the merger hypothesis, appear
now natural in view of our data for the V1309~Sco progenitor.
We do not claim that all observed eruptions of the V838~Mon-type are
mergers of contact binaries. There can be different ways leading to stellar
mergers. What the case of V1309~Sco evidently shows is that the
observational appearances of a stellar merger are indeed the same as
those observed in the V838~Mon-type eruptions.

The outburst of V1309 Sco was shorter and less luminous than those of V838~Mon and 
the extragalactic red novae. 
The latter objects attained luminosities of about or above 
$10^6~{\rm L}_\odot$ and their eruptions lasted a few months.
These differences are most likely caused by the masses of merging stars. For 
V838~Mon, an $\sim8~{\rm M}_\odot$ primary was probably involved
\citep{tylsok06} instead of a $\sim1~{\rm M}_\odot$ 
one of V1309~Sco.

 As noted in Sect.~\ref{sect_cb}, the orbital period of the V1309~Sco
progenitor of $\sim$1.4 day is long for the observed population of
contact binaries. From a study of contact binaries
discovered by the OGLE project in a sky region very close to the position of
V1309~Sco, \citet{rucin98} concluded that the W~UMa type sequence sharply
ends at the orbital period of 1.3--1.5~days \citep[see also][]{pacz06}, 
i.e. just at the orbital
period of the V1309~Sco progenitor. This can be a pure coincidence (just one
case observed), but can also indicate that binaries passing through contact
at periods $\ga$1~day are not rare, but that the contact phase in their case 
is relatively short and quickly leads to a merger.

V1309 Sco, an overlooked object \citep[only one research paper published so
far, i.e.][]{mason10}, 
deserves much more attention of the observers and astrophysicists, as do 
the other V838~Mon-type objects. Apart from supernovae, they belong to the most powerful 
stellar cataclysms. As often happens in nature, cataclysms destroy old worlds, 
but also give birth to new ones. What will develop from the stellar mergers?
Fast rotating giants, similar to FK~Com? 
Peculiar stars with circumstellar discs, when new generation planets can be formed? 
To answer these questions, we just have to follow the evolution of 
the V838~Mon-type objects, V1309~Sco in particular.

\acknowledgements{The OGLE project has received funding from the European
Research Council
under the European Community's Seventh Framework Programme
(FP7/2007-2013) / ERC grant agreement no. 246678.
The research reported in this paper has partly been
supported by a grant no. N\,N203\,403939 financed by the Polish Ministery of
Sciences and Higher Education. The authors are very grateful to
W.~Dziembowski and K.~St\c{e}pie\'n for their valuable comments and
suggestions when discussing various scenarios interpreting the light curve
of the V1309~Sco progenitor.
The periodogram analysis described in this
paper was done using a ZUZA code (version 1.2/rev5) of Alex
Schwarzenberg-Czerny. This paper also uses observations made at the South African
Astronomical Observatory (SAAO). Thanks to the annonymous referee for
the comments, which allowed us to significantly improve the quality of the paper.}
   
\bibliographystyle{aa}

\begin{appendix}
\section{Periodograms from the preoutburst observations}
\label{periodograms}

\begin{figure}
  \includegraphics[scale=0.4]{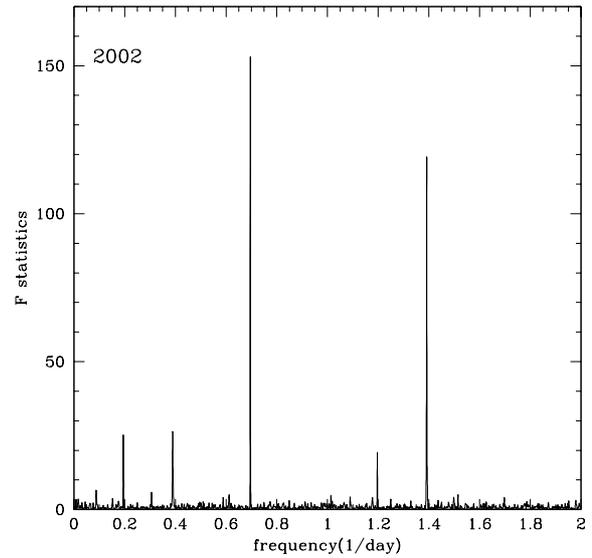}
  \includegraphics[scale=0.4]{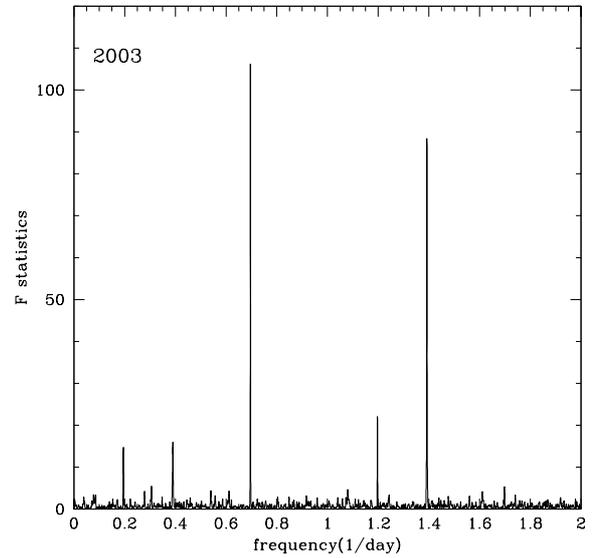}
  \includegraphics[scale=0.4]{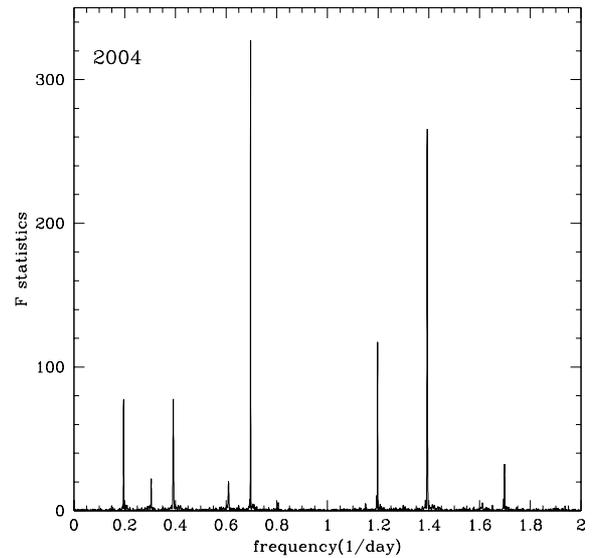}
  \caption{Periodograms from the observations of the progenitor of
V1309~Sco in 2002--2004.}
  \label{per_fig1}
\end{figure}

\begin{figure}
  \includegraphics[scale=0.4]{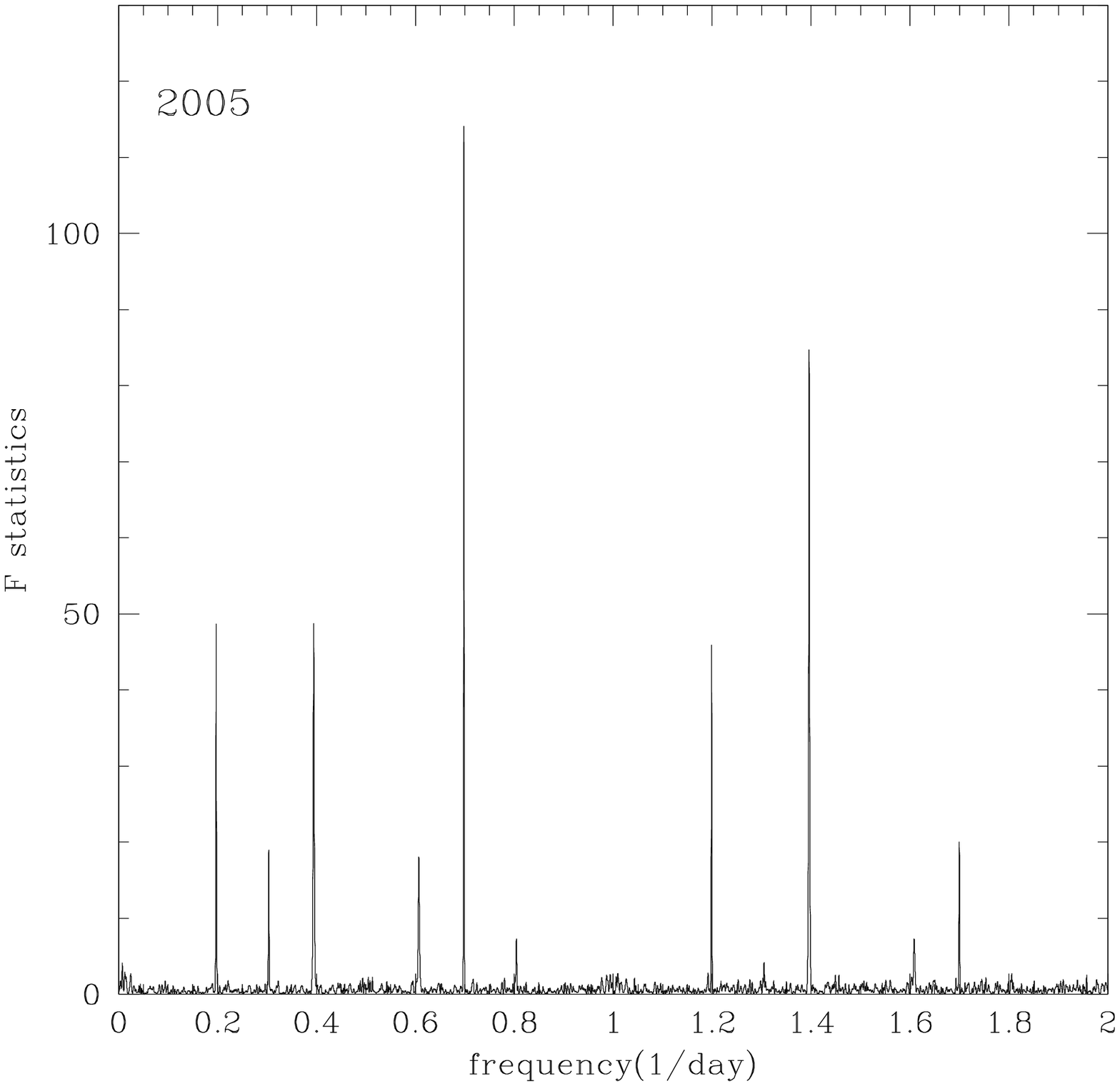}
  \includegraphics[scale=0.4]{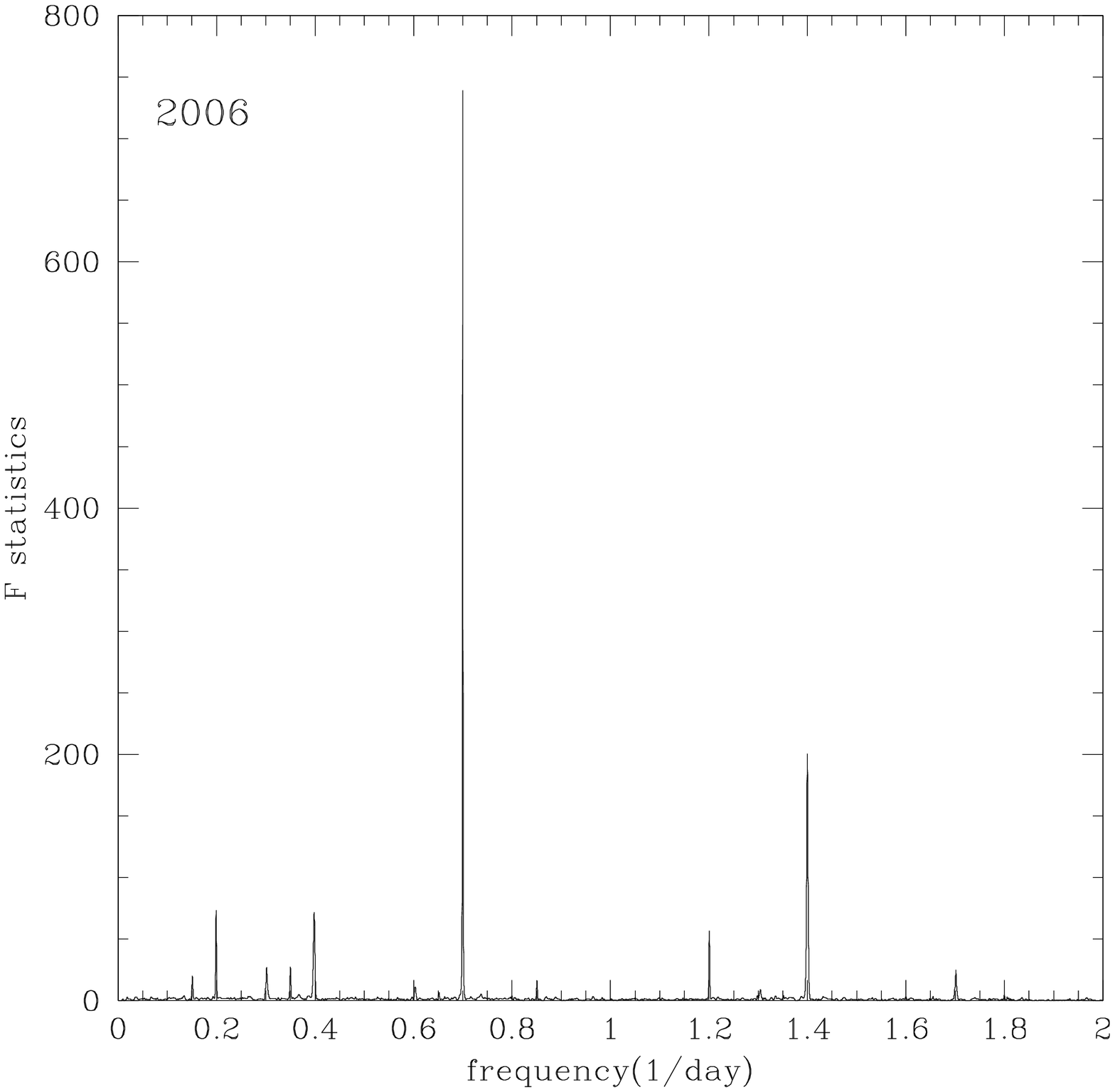}
  \includegraphics[scale=0.4]{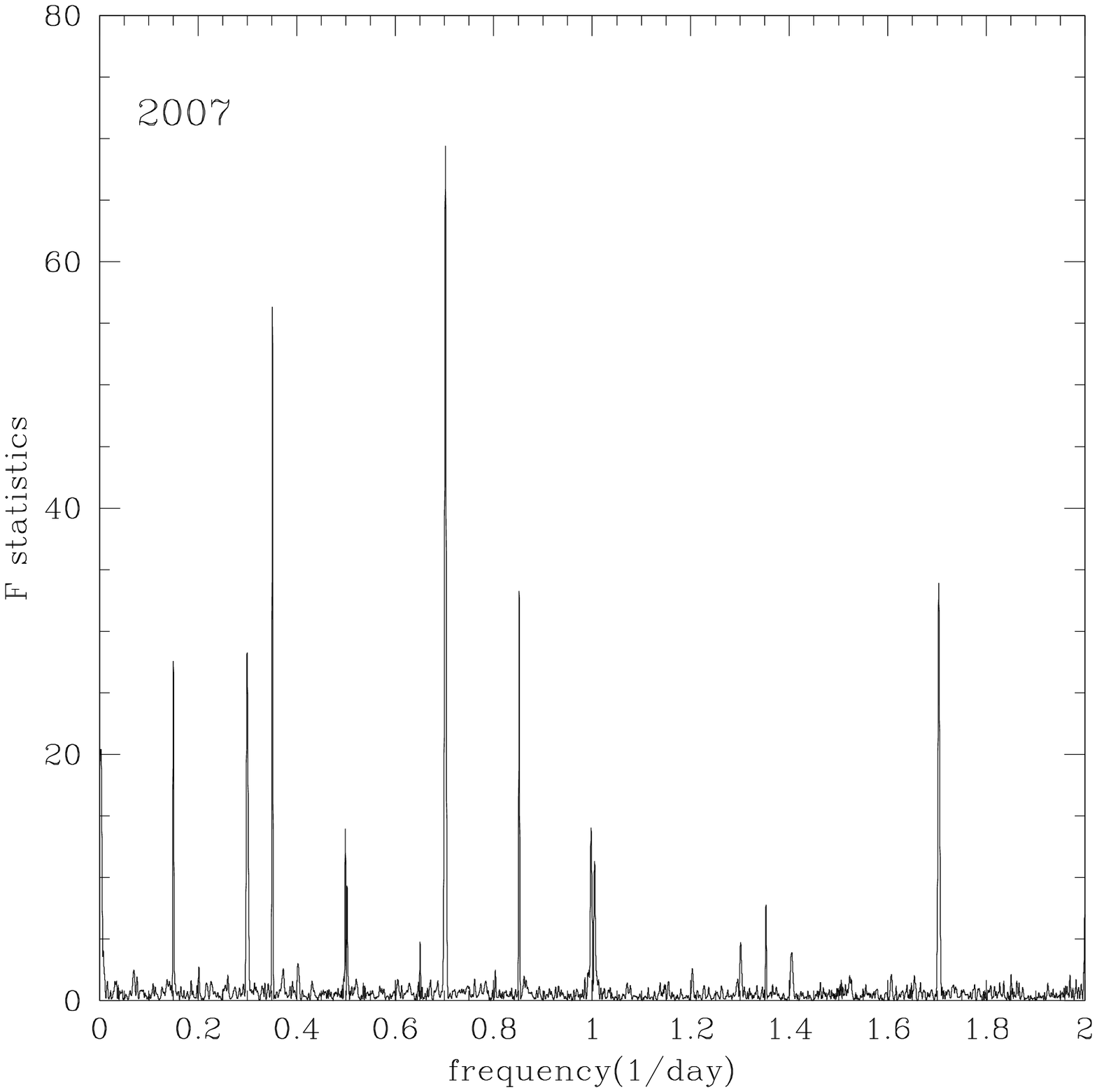}
  \caption{Periodograms from the observations of the progenitor of
V1309~Sco in 2005--2007.}
  \label{per_fig2}
\end{figure}

Figures \ref{per_fig1} and \ref{per_fig2} present periodograms derived from the OGLE
observations of the progenitor of V1309~Sco in 2002 -- 2007.
The periodograms were obtained using the method of \citet{schwarz}, which
fits periodic orthogonal polynomians to the observations and evaluates the
quality of the fit with an analysis of variance (AOV). The resulting AOV statistics
is plotted in the ordinate of the figures.

In all seasons the dominating peaks is at a frequency of
$\sim$0.7~day$^{-1}$, which we interpret as a frequency of the orbital period
of the contact binary progenitor (Sect.~\ref{interpret}).
In earlier seasons, there is also a strong peak at a frequency of
$\sim$1.4~day$^{-1}$, which is because the observations can
also be reasonably fitted with quasi-sinusoidal variations having a period
twice as short as the orbital period.
In 2006 this peak is much weaker and practically disappears in
2007, which reflects the evolution of the light curve as displayed in
Fig.~\ref{fig_lc}. All other peaks are aliases resulting from
combinations of the two above frequencies and a frequency of 1~day$^{-1}$ or
subharmonics of the main peaks.

\end{appendix}

\end{document}